***Proton Irradiation of Primitive Atmospheres of Young Exoplanets and early Earth:***

***$N_2O$ Greenhouse Warming and Prebiotic Synthesis***

Kensei Kobayashi[1,2], Vladimir S. Airapetian[3,4,5,*], Takumi Udo[1], Shunsuke Mouri[1], Yoko Kebukawa[2], Hitoshi Fukuda[6], Yoshiyuki Oguri[6,7], Naoto Hagura[8], M.J. Way[9,10], Guillaume Gronoff[11], Eric T. Wolf[12]

[1]Department of Chemistry and Life Science, Graduate School of Engineering Science, Yokohama National University, 79-5 Tokiwadai, Hodogaya-ku, Yokohama 240-8501, Japan
[2]Department of Earth and Planetary Sciences, Institute of Science Tokyo, 2-12-1 Ookayama, Meguro-ku, Tokyo 152-8551, Japan
[3]Heliophysics Science Division and SEEC, NASA Goddard Space Flight Center, Greenbelt, MD 2077
[4]Department of Physics, American University, Washington, DC, 20016
[5]Graduate School of Advanced Integrated Studies in Human Survivability, Kyoto University, Kyoto, Japan
[6]Research Laboratory for Nuclear Reactors, Institute of Science Tokyo, Meguro-ku, Tokyo 152-8550, Japan
[7]NAT Research Center, Tokai-mura, Naka-gun, Ibaraki 319-1112, Japan
[8]Department of Nuclear Safety Engineering, Tokyo City University, 1-28-1, Takatsuki, Setagaya-ku, Tokyo 158-8557, Japan
[9]NASA/GISS, 2880 Broadway, New York, New York 10025, USA
[10]Theoretical Astrophysics, Department of Physics & Astronomy, Uppsala University, Uppsala, SE-75120, Sweden
[11]SSAI/NASA Langley Research Center, Hampton, VA, USA
[12]University of Colorado, Boulder, Laboratory for Atmospheric and Space Physics, Department of Atmospheric and Oceanic Sciences, Boulder, CO, USA

[*]To whom correspondence should be addressed. Email: vladimir.airapetian@nasa.gov.

**Abstract**
The emergence of habitable conditions on the early Earth and on rocky exoplanets requires persistent energy sources that can drive both prebiotic chemistry and climate warming under magnetically active young G–M stars. To quantify the contribution of stellar energetic particle (StEP) events associated with superflares to the atmospheric chemistry of young planets with primitive atmospheres, we carried out a suite of laboratory proton-irradiation experiments on mildly reduced gas mixtures. We present first proton irradiation experiments of $N_2$–$CO_2$–rich gas mixtures that yield abundant nitrous oxide ($N_2O$) at mixing ratios up to $\sim 10^3$ ppmv, together with amino acid precursors including glycine, corresponding to global production rates of order $2 \times 10^{10}$ kg yr$^{-1}$ on the early Earth. Our photochemical modeling of StEP-driven proton irradiation reproduces the experimentally inferred $N_2O$ production rates and provides self-consistent atmospheric $N_2O$ profiles. We then use these profiles of $N_2O$ as input to a 3D global climate model to evaluate the radiative–climatic impact of StEP-generated $N_2O$ in primitive atmospheres representative of the early Earth and young rocky exoplanets. Our results show that frequent StEP events can help alleviate the faint young Sun paradox on the early Earth and can maintain temperate surface conditions on young rocky exoplanets beyond the outer edges of habitable zone, while simultaneously enhancing the buildup of prebiotic molecules. Together, these processes may constitute a robust pathway toward early planetary habitability.

## 1. Introduction

Understanding how temperate climates and biological precursor molecules emerged in the primitive atmospheres of the early Earth and young rocky exoplanets is a central problem in astrophysics and astrobiology, with direct implications for identifying habitable worlds beyond the Solar System. Addressing this problem requires determining how the key requirements for life as we know it—liquid water, organic compounds, and sustained external energy sources—interact with planetary environments and stellar evolution. Important clues on the initiation of habitable conditions and eventually life forms come from Hadean zircons that show Earth had solid crust and liquid water very soon after formation, and thus life's building blocks formed as early as 4.4 billion years ago (e.g. Wilde et al. 2001). At that time our Sun was a magnetically active star, and, according to the studies of other cooler dwarfs (G to M dwarfs) of similar age, exhibited enhanced magnetic activity characterized by X-ray and EUV emitting hot coronae, and dense magnetized winds (Güdel 2007; Engle 2024). Observations also revealed that both young solar analogs and many cool K–M dwarf frequently produce superflares with energies of ~ $10^{33}$–$10^{35}$ erg (Maehara et al. 2012; Airapetian et al. 2020). Such explosive events are often accompanied by energetic coronal mass ejections (CMEs), giant (masses over $10^{16}$g) and fast (up to 5000 km/s) magnetized clouds with masses over $10^{16}$g detected with Doppler shifts of spectral lines, coronal dimming and Type II events (Namekata et al. 2022; 2024a,b; 2025; Veronig et al. 2025; Callingham et al. 2025). Fast CMEs develop strong shocks in extended coronae of active stars that serve as efficient sites of particle acceleration up to energies of 40 GeV providing the flux of high-energy particles (StEPs) into planetary atmospheres (Hu et al. 2022). Modeling studies have linked StEPs, cosmic rays, and associated space weather processes to nitrogen fixation, organic synthesis, and long-term atmospheric evolution in terrestrial planets' oxygenic atmospheres (Segura et al. 2010; Airapetian et al. 2016; Tilley et al. 2019; Chen et al. 2021; Rigway et al. 2023). These studies suggest the efficient production of NOx molecules that can destroy ozone in Earth-like oxygenic atmospheres via flare events and can promote formation of hydrogen cyanide in anoxic atmospheres of early Earth and young exoplanets. High-fluence energetic protons at energies > 300 MeV associated with superflares can penetrate the lower atmosphere, ionize and dissociate atmospheric nitrogen and other atmospheric species forming free radicals including $NO_x$, NH and $OH_x$. These molecules may serve as efficient sources of HCN, other precursors of life and on early Earth nitrous oxide, a strong "greenhouse" gas on early Earth that can potentially resolve the Faint Young Sun (FYS) paradox, one of the longstanding puzzles of the drivers of early Earth's warm climate [Gough 1982; Felner 2012]. This paradox arises from the geological evidence through Hadean zircons suggesting a temperate climate on early Earth [Wilde et al. 2001], while less than 25% lower solar output (with a modern Earth atmosphere) could have challenged climate conditions supporting the presence of standing bodies of liquid water on its surface [Zahnle et al. 2007]. Later, Kobayashi et al. 2023 (K23) experimentally shown that the production rates of amino acids and carboxylic acids in mildly reducing gas mixtures driven by proton irradiation can significantly exceed the production rates of these molecules via cosmic rays and spark discharges.

Young rocky exoplanets around active G–M stars, analogous atmospheric compositions are expected to develop mildly reducing atmospheres with the primary inventories set by $CO_2$–$N_2$–$H_2O$, with trace $H_2$ and CO, plus sulfur gases such as $SO_2$ and $H_2S$ depending on volcanic activity. These conditions are expected if planetary mantles outgas with similar redox states and if hydrogen escape has not fully oxidized the surface–atmosphere system (Catling and Zahnle 2020).

When applied to diverse exoplanets driven by continuous supply of volcanic or impact-delivered reducing power, different initial conditions and the supply of nitrogen could affect the formation of organic molecules (Bower et al. 2025 and Chen et al. 2025).
Here, we for the first time present the results of laboratory experiments of proton irradiation of diverse mildly reducing gas mixtures, discuss relevant photochemical modeling designed to quantify the chemical products and study their impact on climates of young exoplanets. Section 2 describes the irradiation experiments and resulting production efficiencies of amino acids and $N_2O$, linking stellar magnetic activity to the prebiotic chemical inventory. Section 3 presents the results of 1D photochemical and 3D global climate modeling. In Section 4, we present discussion and conclusions.

## 2. Laboratory Experiments

To cover a wide range of chemical conditions that can be potentially relevant to gas mixtures of primitive atmospheres of young rocky exoplanets and early Earth, we ran two series of proton irradiation experiments at the duration of a StEP ~ 8-10 hours (Hu et al. 2022) of gas mixtures at the total pressure of 700 Torr (0.94 bars) and the room temperature. The first setup used the irradiation of protons at 2.0 MeV from a Tandem accelerator at Tokyo City University, Japan referred to as Case A. The energy of the protons was reduced to 0.94 MeV as they entered the gas chamber, and the beam current was 2 nA, which corresponds to the beam intensity of 1.2 x $10^{10}$ protons/$cm^2$/s with an irradiation time of up to 12 hours. The experimental input proton intensity and duration are comparable with those expected from an individual energetic particle event associated with a superflare based on the size distribution of solar and stellar flares from young solar analogs [Herbst et al. 2019]. A second series of experiments, Case B, were performed by using a Tandem accelerator equipped at Institute of Science Tokyo, where the initial energy of the proton beam at 2.5 MeV was decreased to 1.58 MeV by passing through a Havar foil, with a beam current of 0.5 μA. The proton irradiation time was about 1 h, where the total number of protons was controlled to be 2 mC with an equivalent cumulative energy input from 30 StEP events irradiating an exoplanetary atmosphere in about 100 days [Namekata et al. 2024a,b].
**Case A:** All the samples in the Case A runs initially contained 350 Torr (0.47 bar) of $N_2$ and 5 mL of liquid water. The latter provided about 0.03 bar of water vapor to the gas mixture at ambient temperatures. Sample A1 included 350 Torr (0.46 bar) of $CO_2$ as a carbon species, while Sample A2 had 350 Torr of CO. Sample A3 contained both CO and $CO_2$ (total 0.47 Torr), where the CO mixing ratio was set at 5%. 0.7% of water vapor was added to the system.

**Case B:** Sample B1 refers to $N_2$ (700 Torr or 0.92 bar) and water vapor (ca. 0.03%) in a gas cell, while all other runs were represented by a fixed $N_2$ pressure of 350 Torr (0.46 bar) and variations of methane, carbon dioxide and carbon monoxide with water vapor of $< 1\%$. Sample B2 is represented by equal amounts of methane and $N_2$, Sample B3 contains carbon dioxide in equal (1:1) proportion between methane and carbon dioxide, while Sample B4 replaces carbon dioxide with carbon monoxide in a 1:1 proportion. The Sample B5 contains both CO and $CO_2$ (total 350 Torr), where the CO mixing ratio was set at 8%. 0.7% of water vapor was added to the system. The addition of CO to $CO_2$
The chemical composition of the sample runs is summarized in Table 1.

**Table 1.** *Composition of starting gas phase mixtures of primitive atmospheres*

| **Case #** | **Sample No.** | $N_2$ % | **$CO_2$ %** | **CO** % | **$CH_4$ %** |
|---|---|---|---|---|---|
| Case A | Sample A1 | 50 | 50 | 0 | 0 |
| | Sample A2 | 50 | 0 | 50 | 0 |
| | Sample A3 | 50 | 45 | 5 | 0 |
| Case B | Sample B1 | 50 | 50 | 0 | 0 |
| | Sample B2 | 100 | 0 | 0 | 0 |
| | Sample B3 | 50 | 0 | 0 | 50 |
| | Sample B4 | 50 | 25 | 0 | 25 |
| | Sample B5 | 50 | 42 | 8 | 0 |

### 2.1. Production Rates of Nitrous Oxide Driven by Proton Irradiation.

The experiments show the production of $N_2O$ peaks in all the irradiated products (see the chromatogram Figure 1). To resolve a $N_2O$ peak from $CO_2$ peak with the same *m/z* we used their time history and associate fragments presented in Figure 1. For example, Panel A of Figure 1 shows the total ion chromatogram of Sample B 1 and B2 after 1-hour exposure to proton irradiation. Sample B2 chromatogram shows that $CO_2$ peak appears at $t_R = 3.5$ min, while the $N_2O$ peak occurs at $t_R = 4.025$ min. Also, these products can be discriminated by fragment peaks. For example, the peak at $m/z = 30$ ($NO^+$ fragment) can only be provided by $N_2O$. Changes in the $N_2$ abundance in the gas mixture were kept constant after proton irradiation by using argon as an internal standard. This provided the relative abundances of $N_2O/N_2$ in the irradiation products, thereby allowing us to estimate $N_2O$ production rates.

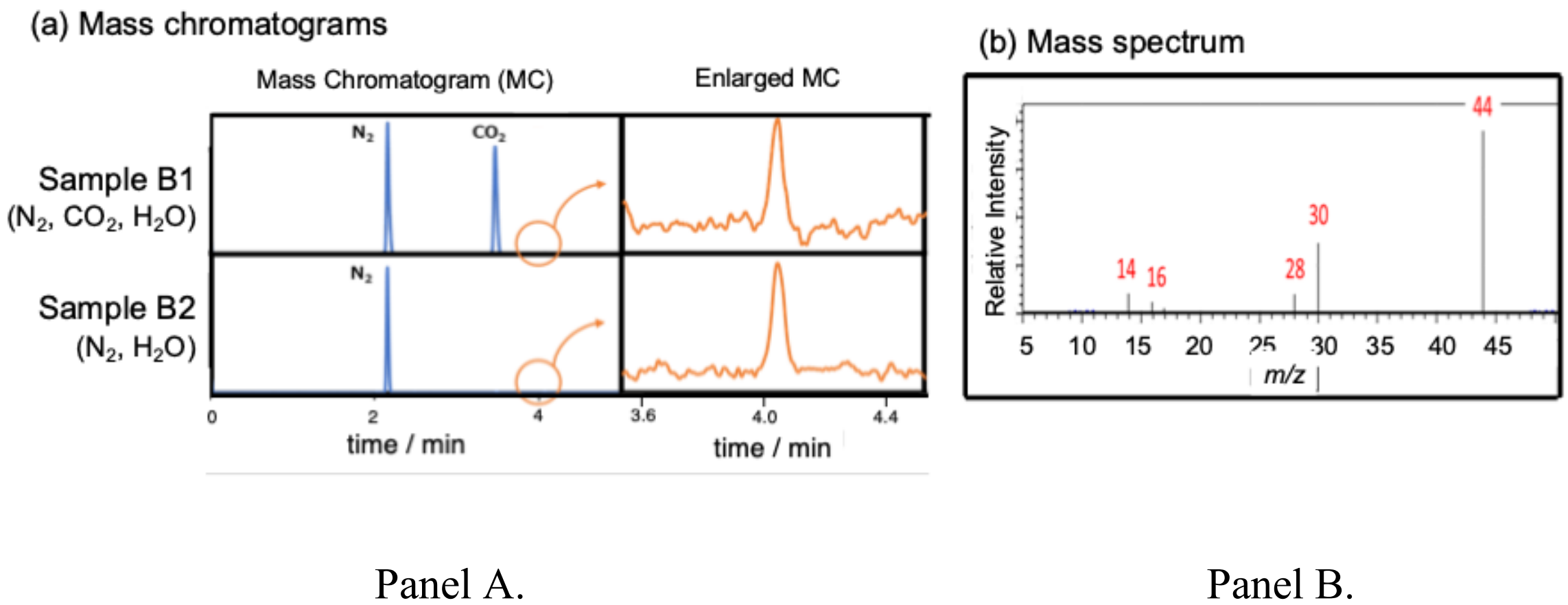

Panel A. Panel B.

**Figure 1**. Panel A: Total Ion Chromatograms of Samples B1 and B2 after proton irradiation; Panel B: Mass spectrum of Sample B2 at $t_R$ = 4.025 min.

Figure 2 shows the $N_2O/N_2$ ratios in the gas mixtures after proton irradiation. At small total input dose (< $10^9$ erg), the $N_2O/N_2$ ratio increased proportionally to the dose as presented in Fig. 2, and it is independent of the mixing ratio of $CO_2$ and/or CO. At the initial $N_2$ pressure of 0.45 bar, the $N_2O$ formation rate (G-value) was 0.4. This suggests the expected rate of 0.02 mol $m^{-2}$ $yr^{-1}$ of $N_2O$ produced during StEP events with the total fluence of $10^{14}$ protons/$cm^2$ per year [Hu et al. 2022]. Table 2 summarizes the equilibrated amount of $N_2O$ produced in Case B experiments. Sample B1 results show the highest production of $N_2O$ among all samples, which suggested that (i) $N_2O$ production depends strongly on $N_2$ concentration, and (ii) $H_2O$ is a good supplier of oxygen atoms. It seems that $CO_2$ also supplied atomic oxygen for $N_2O$ production, since the $N_2O$ formation rate was decreasing in samples as follows: Sample B2 ($CO_2$ - $N_2$) > Sample B5 ($CO/CO_2$ - $N_2$) > Sample B3 ($CH_4$ - $N_2$). Thus, we conclude that not only $H_2O$ but also $CO_2$ and CO could be used as an efficient oxygen source for $N_2O$ production.

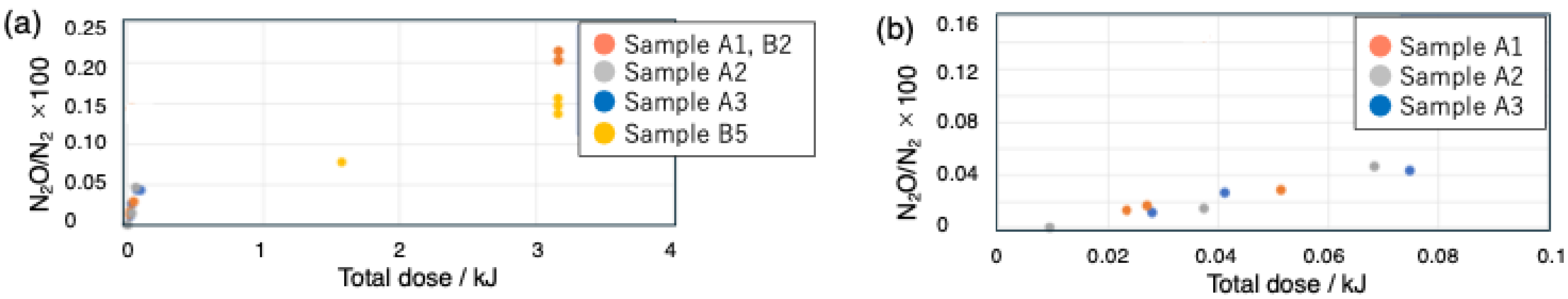

Figure 2. *Panel A: $N_2O/N_2$ ratios in gas mixtures of Sample A1-A3 and B5 for various doses of proton irradiation. Panel B is a zoomed in version Panel A representing the doses between 0 – 0.1 kJ.*

*Table 2. Production of $N_2O$ and glycine in B samples.*

| **Sample No.** | **Initial mixing ratio ($N_2/CO_2/CO/CH_4$)** | **$N_2O/N_2$ (ppmv)** | **Glycine [μmol]** |
|---|---|---|---|
| Sample B1 | **1/1/0/0** | **957±90** | **0.006** |
| Sample B2 | **1/0/0/0** | **1073+100** | **0** |
| Sample B3 | **1/0/0/1** | **541±50** | **15.3[27]** |
| Sample B4 | **1/0.5/0/0.5** | **442±40** | **6.6[20]** |
| Sample B5 | **1/0.84/0.16/0** | **757±70** | **3.24** |

Figure 2 presents the production rates in $N_2O$ normalized to $N_2$ pressure in percents for various doses in Case A (low beam intensity) and Case B (high beam intensity). We can find that the highest production rate of 20% normalized to $N_2$ abundance is found at the total dose of ~ 3.2 kJ for Case A1 and B2 with water vapor and carbon as the sources of oxygen. It is interesting that even at 100 times smaller dose its nitrous oxide production rate with $CO/CO_2$ is about 15%.

### *2.2. Production Rates of Amino Acids*

During proton irradiation experiments, we also analyzed the production of amino acids in Case A and Case B runs using a Shimadzu Amino Acid Analysis System via Cation-exchange HPLC [K23]. Figure 3 presents the HPLC chromatograms of amino acids from the Sample B4 ($CH_4$ - $CO_2$ - $N_2$ - $H_2O$) irradiation run after acid-hydrolysis and shows the formation glycine as the the dominant product followed by aspartic acid, serine, alanine, α-aminobutyric acid and β-alanine (see details in SI). In Sample B1 ($CO_2$ - $N_2$ - $H_2O$), only traces of amino acids (mainly glycine) were detected. Since the irradiation products before acid-hydrolysis yielded only trace level of amino acids, this means that amino acid precursors rather than free amino acids were formed. We detected no aminoacetonitrile ($NH_2CH_2CN$) in the sample without hydrolysis as would be expected because aminoacetonitrile would change to glycine after hydrolysis. This suggests that the Strecker synthesis is not the dominant formation pathway in prebiotic synthesis of amino acids via proton irradiation. The average yield of glycine in the Sample B5 [$N_2$-$CO/CO_2$-$H_2O$] run after hydrolysis was 3.24 μmol. Energy yields in radiation chemistry are usually expressed as G-values, which is the number of molecules produced per 100 eV [Baird et al. 1990; K23]. We obtained G-value for glycine of $1.0 \times 10^{-2}$, which was about half of that obtained from a mixture of CO, $N_2$ and $H_2O$ [Kobayashi et al. 1997]. In the case when $CH_4$ was used as a reducing carbon source in a mixture of $CO_2$, $CH_4$, $N_2$ and $H_2O$, glycine yield decreased nearly linearly with the decrease of $CH_4$ molar ratio [K23]. In the case of Sample B5, glycine yield was only slightly lower even when the CO molar ratio was largely reduced. This suggests that StEPs are an effective energy source to yield amino acid precursors in $N_2$-$CO/CO_2$-rich planetary atmospheres when no methane in the gas mixture was present, and water vapor is an efficient source of hydrogen.

If we assume the energy flux of a StEP event associated with a superflare with the energy of $10^{34}$ erg precipitating on early Earth's atmosphere (at 4.3 Ga) is $3 \times 10^{14}$ protons/cm$^2$/yr at the occurrence rate of 1 event per 3 days [Namekata et al. 2024a], then the annual cumulative production of glycine on early Earth's surface is expected to be $2 \times 10^{10}$ kg yr$^{-1}$. Given that the estimated glycine flux via exogenous delivery rate by carbonaceous chondrites on early Hadean Earth was ~ 1 kg yr$^{-1}$ [Kobayashi et al. 1997], we conclude that endogenous amino acid production rate by StEP events from the young Sun is likely to be significantly higher.

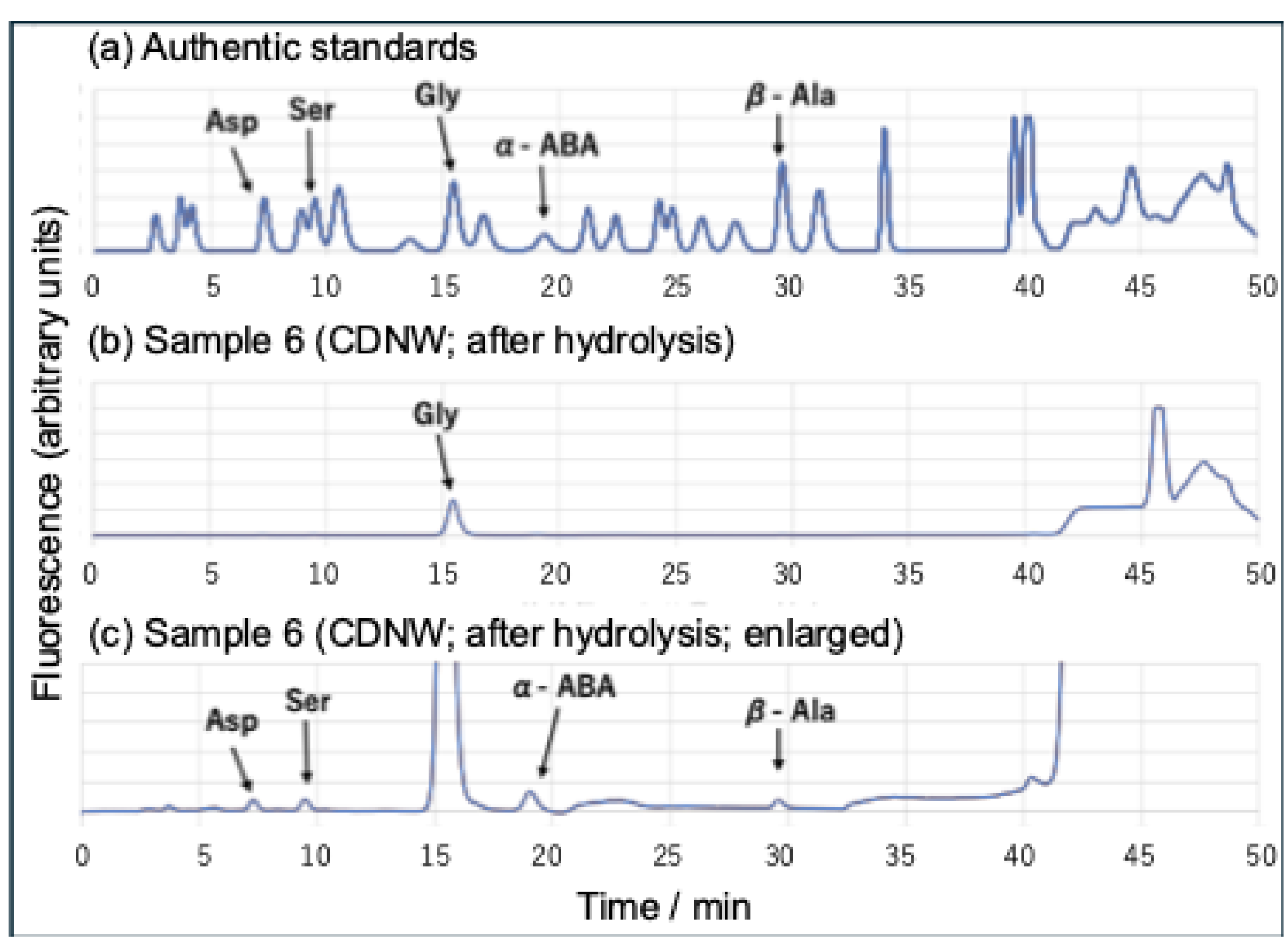

**Figure 3.** The figure shows the HPLC Chromatograms of Amino Acids produced in the Sample B5 irradiation run.

## 3. Photochemical and Climate Modeling of Early Earth and Young Exoplanets

To characterize the $N_2O$ production rate, we applied the Aeroplanets model with an enhanced chemistry [Airapetian et al. 2016]. The model calculates the photo and collisional (due to protons) dissociation, ionization and photoexcitation processes in a gas mixture. The primary photoelectrons are then transported in the chamber via collisions, and the electron impact is computed by solving the stationary kinetic Boltzmann equation. Aeroplanets is coupled with the Planetocosmic model (Gronoff et al. 2009). Planetocosmics computes the transport of relativistic particles in planetary atmosphere using the GEANT-4 model. This results in the dissociation, ionization and excitation of the different atmospheric species. The Aeroplanets code incorporates 117 chemical reactions [Airapetian et al. 2016].

We first simulated the production of $N_2O$ via a monoenergetic 0.9 MeV proton beam irradiation of the 80%$N_2$ /20%$CO_2$ gas mixture at 1 bar, $T=20^{0}C$ in the gas chamber with the length of 20 cm and the width of 6 cm. A steady-state chemical solution for the gas mixture was reached after running the code for 1 week of physical time and produced ~100 ppmv of $N_2O$. The stopping length of the proton beam at 0.9 MeV is ~ 6 cm, which suggests that the contamination of chamber's walls by protons do not affect the experimental results.

We also applied the Aeroplanet model to simulate the propagation and chemical impact of a StEP event associated with a $10^{34}$ erg superflare in a gravitationally stratified 80%$N_2$ /20%$CO_2$ atmosphere. The left Figure 4 shows the StEP energy spectrum associated with a superflare from an active (125 Myr old) solar-like star (M3 model in Hu et al. (2022). As StEPs with the hard energy spectrum propagate into the planetary atmosphere, they ionize and dissociate atmospheric species ($N_2$ and $CO_2$) and create a cascade of secondary electrons with the maximum energy deposition in the upper troposphere shown in the right panel of Figure 4. Energetic electrons produce free radicals including NOx (NO, $NO_2$), OH and NH, the major prerequisite species for formation of $N_2O$ via $N_2 + O \rightarrow NO + N(^2D)$; $N + OH \rightarrow NO + H$; $NO + O \rightarrow NO_2$; $N+NO_2 \rightarrow N_2O + O$; $N+NO_2 \rightarrow NO + NO$ and $NO+NH \rightarrow N_2O + H$.

The left panel of Figure 5 shows the vertical profiles of resulted C and N-bearing species with the $N_2O$ concentration of 400 ppmv (see details in Hayworth 2022), and thus consistent with our experimental results. However, the lower yield of $N_2O$ in the model as compared to Table 2 experimental yields can be explained by the lack of ion-neutral model chemistry via $N_2^+$, $CO_2^+$, $O^+$, $NO^+$.

Our experiments show that $N_2O$ production scales positively with $N_2$ abundance as its main supplier and the availability of O-bearing species ($H_2O$, $CO_2$), while the detailed pathways depend on the relative contributions of NO, $NO_2$, and NHx. If young rocky exoplanets begin with substantially larger or smaller $N_2$ inventories than Earth, as suggested by Bower et al. (2025) and Chen et al. (2025), their steady-state $N_2O$ columns under comparable StEP forcing could be orders of magnitude higher or lower than in our $N_2O$ analog case.

Our experiments study the $N_2O$ production rates in an individual StEP event which occurs at the frequency of one event per 3 days (for a 125 Myr old Sun) or weeks (for a 400 Myr old Sun), which is much shorter than the $N_2O$ residence time of ~ 20 - 40 years estimated by our photochemical AEROPLANETS model [Airapetian et al. 2016]. This means that eruptive events associated with superflares from the young solar analogues may have contributed to the accumulation of $N_2O$ in the young rocky exoplanets and Hadean Earth atmosphere over a few hundred Myrs as nitrous oxide is not very soluble in water, and thus it would not rain out to the planetary surface. Thus, $N_2O$ production via StEPs may have important implications for the climates of young rocky exoplanets and early Earth as nitrous oxide's warming potential is much greater than $CO_2$ (see Section 3).

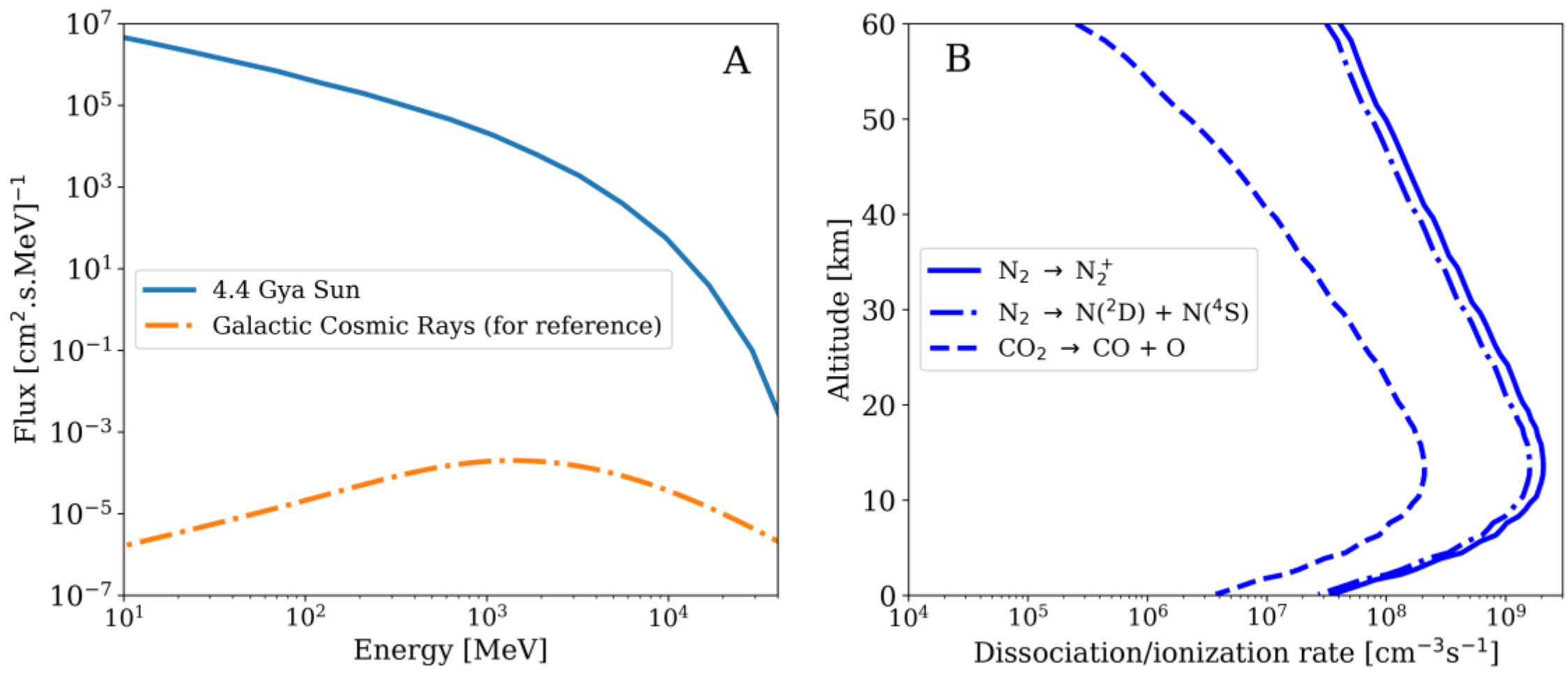

Figure 4. Panel A: Energy spectrum of a StEP associated with a CME event at 3000 km/s (M3 event, Hu et al. 2022); Panel B: Dissociation and ionization rates due to collisions with energetic protons and developed cascade of secondary electrons.

We used the three-dimensional (3D) Resolving Orbital and Climate Keys of Earth and Extraterrestrial Environments with Dynamics (ROCKE-3D version 1.0 [Way et al. 2017]) global climate model (GCM) to study the climate solutions for the early Archean Earth (3.8 Gya, 75% solar luminosity) with an aquaplanet topography and an atmospheric composition of 90%$N_2$/10%$CO_2$/400 ppmv of $N_2O$ to obtain a global mean surface temperature (GMST) averaged over the last 10 years of the model run of $T_{GMST}$ = +34$^0$C. Our recent simulation with 100 ppmv of $N_2O$ concentration yields $T_{GMST}$ = +30$^0$C only 4$^0$C lower than that at 400 ppmv if no atmospheric methane is included, while $T_{GMST}$ = + 18$^0$C without $N_2O$ abundance (a difference of 12$^0$C). However, with only 1% CO2 and 100ppmv $N_2O$ we could obtain a narrow water belt state (Panel B of Figure 5), while increasing $N_2O$ to 400ppmv yielded a ice free state. Thus, our global climate model strongly suggests that $N_2O$ at these and even smaller concentration can present a viable solution for the resolution of FYS paradox under low $CO_2$ mixing ratio and no methane present. For the 3D GCM results the model setup included modern Earth orbital parameters, but with eccentricity=0, and a 24-hour length of day (Lod). It was run for 1000 years until radiative equilibrium was achieved. If one were to use an 18-hour LoD (a typical value given for the early Archean; Williams et al. 2000) the equator to pole meridional temperature gradient is expected to increase (Kienert et al. 2012), but the GMST should not change more than 1–2$^0$C from unpublished work by the ROCKE-3D team. This is small change in GMST is consistent with work by Jenkins et al. 1996; Kienert et al. 2012 did not publish a GMST for their 24hr vs 18hr LoD models.

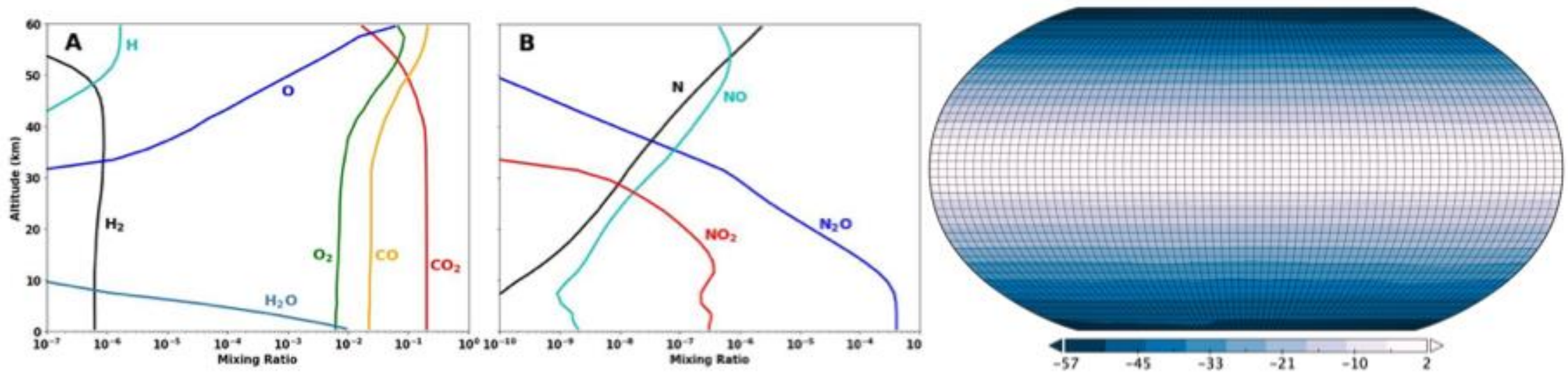

Figure 5. Panel A and B: Production rates of C and N-bearing molecules including $N_2O$; Panel C. The ROCKE-3D model GMST map due to the impact of 100 ppmv of $N_2O$ in the 99%$N_2$1%$CO_2$ atmospheric composition of the early Earth or a rocky exoplanet receiving 75% of the current solar luminosity.

## 4. Discussion and Conclusions

In this study, we assessed the chemical and climate impacts of energetic particle events associated with superflares at $10^{34}$ erg occurred at the rate of 1 event per 3 days. We irradiated three samples of mildly reduced gas mixtures with low (Case A) and five samples with high (Case B) intensity proton beams for a few hours and have shown the efficient production of > 400 ppmv of $N_2O$ and abundant production of amino acids including alanine, glycine and others. These production rates of $N_2O$ exceed the mixing ratios formed via lightning sources and corona discharge, and impact experiments simulating bolide collisions [Nna-Mvondo et al. 2005; Heyas et al. 2002].

The production of 957 ppmv of $N_2O$ in the Sample 2 run at 0.9 bar suggests the abundance of 8 x $10^{12}$ molecules/J of $N_2O$, while the experiments via corona discharge of $CO_2$ (20%) -$N_2$ (80%) gas mixture at 1 bar produced ~ 3 times less $N_2O$ per J. The higher production rate of $N_2O$ per J is caused by the impact ionization, dissociative ionization, excitation and dissociation of $N_2$, $H_2O$ and $CO_2$ via protons, pions, gamma-rays as a part of the whole electromagnetic cascade induced by a simulated StEP event. Large fluxes of energetic electrons driven by auroral electron precipitation and CME events were observed in the wake of the 2003 CME events show the production of $N_2O$ together with high abundance of $NO_x$ in Earth's atmosphere [Semeniuk et al. 2008; Funke et al. 2008].

In addition, our experimental study sheds light on the pathways of production of amino acids in relation to the enhancement of $N_2O$ from a range of mildly reduced molecular mixtures reflecting possible primitive atmospheres of rocky exoplanets and early Earth. We observed the production of glycine, formation of other amino acids including that in C-containing molecules in Sample B runs. This suggests that the production of precursor molecules depends on the starting chemistry. We conclude that that the production rate of glycine in gas-phase experiments is i) nearly dependent on the starting mixing ratio of methane; ii) nearly linearly dependent on the input energy; iii) can be moderated by water vapor in the absence of methane at high radiation doses (3.2 kJ). This experimentally derived conclusions can be understood from the efficient production of highly reactive atomic hydrogen rich in methane or water vapor (conclusion i and ii). Conclusion iii) can be understood in terms of a large population of nonthermal particles efficiently dissociating water vapor into OH and H. One of the major results of our study is the experimentally obtained correlation between G-values of amino acids and G-values of nitrous oxide for Samples B1-B5 (see Table 2). We have also shown that glycine and other amino acids can be formed with only 0.7% of water vapor serving as the efficient source of hydrogen (Sample 5B run), and thus no methane is required for efficient production of this precursor molecules if their formation is mediated by a proton irradiation event. An individual StEP event can generate over 700 ppmv of $N_2O$ and $2 \times 10^{10}$ kg yr$^{-1}$ of glycine in such a mildly reduced atmosphere. This suggests an efficient accumulation of $N_2O$ in the atmosphere and glycine at the surface given the high frequency of occurrence of StEP events over 200-300 million years. These results are consistent with the expectations of our photochemical model.

$N_2O$ is also considered to be a nitrogen source for synthesizing ammonia and organic nitrogen species including amino acids under UV irradiation from the young Sun [Zang et al. 2022]. This synthesis would have served as a great impetus for the initiation of life on early Earth in the first few hundred Myr and rocky exoplanets around young, active K-M type dwarfs. As our 3D GCM results strongly suggests that $N_2O$ at these and even smaller concentration (100 ppm) can present a viable solution for the resolution of FYS paradox at 1% of $CO_2$, they can potentially resolve the FYS paradox for a low $CO_2$ atmosphere on early Earth, suggest a pathway for initiation of prebiotic chemistry and expand the habitable zones of rocky exoplanets around active stars.

A recent study demonstrates that the $N_2O$ concentration at 1000 ppmv can be potentially detectable in an Earth-like reflection spectra by HWO (Tokajian et al. 2024). Both prebiotic organics and strongly absorbing greenhouse gases currently under further studies at Exoplanetary Particle Irradiation Chemistry (EPIC) lab at GSFC—may produce observable spectral signatures accessible to next-generation observatories such as the Habitable Worlds Observatory.

### *Acknowledgments and funding sources*

The authors wish to thank the anonymous referee for the constructive comments and suggestions that improved the quality of the paper. The authors also acknowledge Dr. Yoshihiro Kubota and Dr. Satoshi Inagaki (Yokohama National University) for their kind help in GC/MS analysis, and Dr. Jun-ichi Takahashi and Dr. Hiromi Shibata (Kobe University) for their useful discussion on irradiation experiments and chemistry of early Earth. This study was in part supported by the Japan Society for the Promotion of Science (JSPS) through grant-in-aid Nos. 20H02014, 23K03561, and 23H01286. Vladimir S. Airapetian (VSA) acknowledges support from the NASA/GSFC Internal Research and Development funds (IRAD) to develop EPIC lab at GSFC and Sellers Exoplanet Environments Collaboration (SEEC), the Fundamental Laboratory Research (FLaRe), which are funded by the NASA Heliophysics and Planetary Science Division's Internal Scientist Funding Model (ISFM) and NASA's Astrophysics Theory Programs program. Vladimir S. Airapetian and Kensei Kobayashi acknowledge the International Space Science Institute and the supported International Team 464. This work was supported by NASA's Nexus for Exoplanet System Science (NExSS) and the NASA Interdisciplinary Consortia for Astrobiology Research (ICAR). Resources supporting this work were provided by the NASA High-End Computing Program through the NASA Center for Climate Simulation at Goddard Space Flight Center. G. Gronoff's work was supported by the NASA SMD, Heliophysics Division and Space Weather Science Applications Program. M.J.W. acknowledges support from the GSFC Sellers Exoplanet Environments Collaboration (SEEC), which is funded by the NASA Planetary Science Division's Internal Scientist Funding Model, and ROCKE-3D, which is funded by the NASA Planetary and Earth Science Divisions Internal Scientist Funding Model.

**Author Contributions**: Conceptualization and original draft preparation, K.K. (Kensei Kobayashi), V.S.A (Vladimir S. Airapetian); conducting experiments and analysis, T.U., S.M. and Y.K.; interpretation of the data, K.K., Y.K., V.S.A; conducting proton irradiation, H.F., Y.O. and N.H.; performing simulations, Guillaume Gronnof, Michael Way, Eric Wolf, V.S.A.; funding acquisition. All authors have read and agreed to the published version of the manuscript.

## **Supplementary Materials**.

## S1. Methods and Protocols

### *S1-1. Chemicals*

Pure $CO_2$ and $N_2$ gases were purchased from Suzuki Shokan Co., Japan. Pure $N_2O$ gas was purchased from Koike Medical Co., Japan. A 50.0: 50.0 (v/v) gas mixture of $CO_2$ and CO was prepared by Taiyo Nippon Sanso Co., Japan. $^{13}$C-labelled CO was purchased from Shoko Science Co., Ltd. ($^{13}C > 99\%$).The followings were used as standards of Amino acids: Amino acid standard mixtures Type AN-II and B. HCl (amino acid analysis grade, Fujifilm Wako, Japan) was used to hydrolyze irradiation products. These reagents were purchased from Fujifilm Wako Pure Chemical Co.,

Japan. o-Phthalaldehyde and N-acethyl-L-cystein used for post-column derivatization of amino acids were purchased from Fujifilm Wako Pure Chemical Co., Japan. 2,2,3,3,4,4,4-heptafluoro-1-butanol (Tokyo Chemical Industry, Japan), ethyl chloroformate and chloroform (Fujifilm Wako Pure Chemical Co., Japan) were used for derivatization of amino acids for GC/MS analysis. All the glassware and metals used were heated in an electric oven at 500°C to remove organic contaminants. Water used was purified with a Milli-Q system (Merck KGaA, Germany).

*S1-2. Instruments*

Cation-exchange HPLC used was a Shimadzu LC-10AT Amino Acid Analysis System with a RF-20Axs fluorescence detector, and a Shimpak ISC-07/S 1504 column (4.0 mm i.d. × 150 mm), where amino acids in the effluent were derivatized with o-phthalaldehyde and N-acetyl-L-cysteine for fluorometric detection. Amino acids were also analyzed by gas chromatography-mass spectrometry (GC/MS; a Shimadzu GCMS-QP2020; with an Agilent CP-Chirasil Val column (25 m long × 0.25 mm i.d. × 0.12 mm film thickness) after derivatization.

*S1-3. Determination of gaseous products*

Gaseous products were analyzed by GC/MS (Shimadzu GCMS-QP2000) with an Agilent J&W GC column PoraPLOT-Q (0.25 mm i.d. × 25.0 m, film thickness: 8 mm) with a spilt injection (1/100) and an inlet pressure of 51.2 kPa. Carrier gas used was helium. Two kinds of oven temperature programs were applied: (i) hold at 30 °C for 4 min, increased up to 90°C at 4 °C/min, and then increased up to 200 °C at 20 °C/min; (ii) hold at -50 °C with a low-temperature control solenoid valve set CRG-2030 for 4 min, increased up to 90°C at 4 °C/min, and then increased up to 200 °C at 20 °C/min. The injection port temperature was 250 °C. The program (ii) was used to determine lower level of $N_2O$. Electron ionization (EI) was applied at 70 eV with the ion source temperature of 200 °C and the scanning *m/z* range of 10–150. The following *m/z*'s were used to determine the concentration of the products: $CH_4$ (*m/z* 16), CO (*m/z* 12, 28), $N_2$ (*m/z* 14, 28), $CO_2$ (*m/z* 44), Ar (*m/z* 40) and $N_2O$ (*m/z* 44).
$N_2O$ was also detected in all the irradiated products. Formation of the other N-containing species, such as HCN and $NO_2$ was theoretically expected in dry air [Kobayashi et al. 1998], but their peaks were not identified because HCN is easy to dissolve in water, and $NO_2$ is highly reactive with water to form $HNO_3$, which would be the reason why they could not be detected in the gas phase products.

*S1-4. Analysis of amino acids*

Amino acids were determined by Ion-exchange high performance liquid chromatography (HPLC; Shimadzu LC-10AT) after acid hydrolysis at 110 °C for 24 h in 6 M HCl. Analysis details were presented in [Kebukawa et al. 2022]. Amino acids were also analyzed by GC/MS after N-ethoxycarbpnyl heptafluorobutyl ester derivatization (Ubukata et al., *JMSSJ*, 2007). The derivatives were injected to the gas chromatograph-mass spectrometer. Derivatization procedures and analytical conditions are described in [Taniuchi et al. 2013.]. Figure S2 shows the HPLC chromatograms of amino acids produced in the Sample B5 irradiation run.